# An Efficient Transparent Test Scheme for Embedded Word-Oriented Memories

Jin-Fu Li, Tsu-Wei Tseng, and Chin-Long Wey
Advanced Reliable Systems (ARES) Laboratory
Department of Electrical Engineering
National Central University
Jungli, Taiwan 320, R.O.C.

**Abstract**

*Memory cores are usually the densest portion with the smallest feature size in system-on-chip (SOC) designs. The reliability of memory cores thus has heavy impact on the reliability of SOCs. Transparent test is one of useful technique for improving the reliability of memories during life time. This paper presents a systematic algorithm used for transforming a bit-oriented march test into a transparent word-oriented march test. The transformed transparent march test has shorter test complexity compared with that proposed in the previous works [12, 13]. For example, if a memory with 32-bit words is tested with March C−, time complexity of the transparent word-oriented test transformed by the proposed scheme is only about 56% or 19% time complexity of the transparent word-oriented test converted by the scheme reported in [12] or [13], respectively.*

## 1  Introduction

Shrinking transistor size makes the reliability issue become a major challenge of system-on-chip (SOC) designs. Nowadays SOCs usually consists of many memory cores, which are usually the densest portion with the smallest feature size. Thus the reliability of memory cores has heavy impact on the reliability of SOCs. Reliability enhancement techniques for memory cores during life time thus is imperative. Conventional memory BIST (off-line BIST) is a promising approach for embedded memory testing and diagnosis [1, 7, 10, 16], which is helpful for improving the yield of memories during manufacturing phase. However, off-line BISTs cannot be used for testing memories during life time. Thus efficient reliability enhancement techniques for memories should be developed. Online and transparent (periodic) testing are two widely used methodologies for ensuring correct operation of memories during life time.

Concept of transparent testing, which leaves the original contents of the circuit under test unchanged after the testing is completed, has been used for the application of memory testing [3, 6, 8, 9, 11–13, 17–19]. One major advantage of transparent testing is that it can ensure the reliability of storage data during a life-time operation. On the other hand, transparent testing provides better fault coverage than non-transparent testing for unmodeled faults. Thus, unmodeled faults not detected by the manufacturing testing will be discovered at an early stage of the product life [12].

Several transparent test schemes have been reported in [3, 6, 8, 9, 11–13, 17–19]. These schemes transform the march tests, which have been widely used to test random access memories, into transparent march tests. In [11, 12], a systematic approach for transforming a march test into a transparent march test is presented. The transformation rules consist of two phases—generation of transparent march test and generation of signature prediction test. In [3, 8, 17], the method from [11] is applied to the tests for detecting pattern-sensitive faults and single V-coupling faults described in [2]. Later a symmetric transparent test methodology is proposed in [18]. In this methodology, if the transformed transparent test is not symmetric, then an additional state is added to make it be symmetric. This causes that the final content of the signature analyzer (e.g., MISR) is zero if no faults exist in the memory under test. Thus this methodology can reduce the test time by removing the signature prediction test. Automatic generation of symmetric transparent march tests is also proposed in [19]. However, the transparent tests described above all have the problem of aliasing.

Recently, transparent test schemes without the aliasing problem have been reported in [9, 13]. In [9], a transparent test approach using dynamic power supply current (DPSC) is presented. Instead of generating a signature, the DPSC transparent testing detects the RAM faults by using a current sensor. A transparent online memory test (TOMT) reported in [13] has been developed for online testing of word-oriented memories with parity or Hamming protection. TOMT introduces the concept of concurrent error detection and correction to eliminate the requirement of signature generation and to avoid the interference of normal system operation. So far, however, all the transparent tests for word-oriented memory are not time efficiency. For example, the transparent test scheme reported in [11, 12] performs the bit-oriented operations to all the bits of each word. Another example, TOMT also executes bit-wise ma-nipulations in word-oriented memory testing. Therefore, short transparent tests for word-oriented memories need to be developed. Moreover, shorter test time can reduce the probability of interference of normal system operation, since transparent tests usually are executed in idle state of systems.

This paper presents an efficient transparent test scheme for word-oriented memories. A systematic algorithm is used for transforming a bit-oriented march test into a transparent word-oriented march test. Consider an $N \times B$-bit memory and a bit-oriented march test with P Read/Write operations in which there are Q Read operations. Time complexity of the transparent word-oriented march test converted by the proposed algorithm is $(P+5\log_2 B)N$. Also, time complexity of the corresponding signature prediction test is $(Q+2\log_2 B)N$. The transformed transparent march test has shorter test time compared with that proposed in previous works. For example, if a memory with 32-bit words is tested with March C−, the time complexity of the transparent test transformed by the proposed scheme is only about 56% or 19% time complexity of the transparent test converted by the scheme reported in [12] or [13].

The rest of this paper is organized as follows. Section 2 reviews typical functional RAM faults and defines the notation for representing algorithms. Section 3 describes the transformation rules for conventional transparent march tests. Section 4 introduces the proposed transformation algorithm for transparent word-oriented march tests. Section 5 shows the fault coverage analysis of the transparent word-oriented march tests transformed by the proposed transformation algorithm. Section 6 concludes this paper.

## 2 Fault Model and Notation

In this section we present the widely used fault models of RAMs, which consist of stuck-at faults, transition faults, and coupling faults [4,5,14]. More detail behaviors of these faults are described as follows.

1. Stuck-at fault (SAF)—the defective cell permanently contains a 0, i.e., SAF(0), or a 1, i.e., SAF(1), and cannot be changed.

2. Transition fault (TF)—the defective cell fails to undergo a 0 to 1 transition or a 1 to 0 transition.

3. Coupling fault (CF)—the status of a cell (called coupling cell) affects the content of the other cell (called coupled cell). CFs can be divided into three types: state CF (CFst)—the coupled cell is forced to 0 or 1 only if the coupling cell contains 0 or 1; idempotent CF (CFid)—the content of victim is forced to 0 or 1 if the aggressor undergoes a 0 to 1 transition or a 1 to 0 transition; inversion CF (CFin)—the content of victim is inverted if the aggressor undergoes a 0 to 1 transition or a 1 to 0 transition. In word-oriented memories, CFs can further be divided into intra-word and inter-word

CFs. Intra-word CFs are defined as that the coupled and coupling cells of the CFs are within a word; inter-word CFs are defined as that the coupled and coupling cells of the CFs are located at different words.

Subsequently, notation for algorithm representation is described. In an algorithm description, $D$ denotes the initial content of a cell or a word for bit-oriented or word-oriented memories, respectively; $D_a$ is the data $D \oplus a$, where $\oplus$ denotes bit-wise XOR operation, where $D$ and $a$ have the same data width; $\Uparrow$ ($\Downarrow$) represents the ascending (descending) address sequence; $\Updownarrow$ denotes either ascending or descending address sequence. $wX$ denotes write $X$ operation to an addressed memory cell (word). $rX$ denotes read operation from an addressed memory cell (word) and $X$ is expected. In an nontransparent march test, the $X$ is a specific data for Write operation. On the other hand, the $X = D_a$ for Write operation in a transparent march test.

## 3 Conventional Transparent March Tests

March tests (algorithms) are widely used for memory fault detection and diagnosis due to their linear time complexity with respect to the memory size. A march test consists of a finite sequence of march elements. Each march element contains a finite number of Read/Write operations to all cells (words) according to a prespecified address sequence of bit-oriented (word-oriented) memories.

Performing march tests changes initial contents of the memory under test. Thus march tests cannot be used to test the memories in power-up status, i.e., march tests cannot directly be used for online test. Online tests can be divided into concurrent and nonconcurrent online tests. Concurrent online memory testing usually introduces error detection and correction techniques to preserve the data integrity of the memory under test. Nonconcurrent online memory testing usually uses transparent tests to detect the faults and to preserve the data integrity by writing back the read data when the memory under test is in idle state. Thus transparent march tests could be used for periodic online memory testing. This allows restoring memory content after transparent march test procedure.

A conventional bit-oriented march test can be converted into a bit-oriented transparent march test by the following transformation rules [11, 12]:
**Step 1:** If the first test operation of a march element is a Write one, add a Read operation at the beginning of the march element. Subsequently, if a march test consists of an initialization march element and if the march element is useless for fault activation, then remove the march element from the march test.
**Step 2:** Replace all $r0$ or $r1$ operations with $rD_a$ or $rD_{\bar{a}}$, respectively. Also, replace all $w0$ or $w1$ operations with $wD_a$ or $wD_{\bar{a}}$ operations, respectively, where $a=0$.
**Step 3:** If the content in memory cells are the inverse of their initial data after the last Write operation, then insert two additional Read and Write operations in the end of the march test. The additional two operations perform Read operation on each memory cell followed by a Write operation with the data which is the inverse of the read data.

**Step 4:** Remove all the Write operations from the transparent march test, which is converted according to Step 1 to Step 3, to obtain the signature prediction algorithm.

For example, consider the classical March C− test [14]: $\{\updownarrow (w0); \Uparrow (r0, w1); \Uparrow (r1, w0); \Downarrow (r0, w1); \Downarrow (r1, w0); \updownarrow (r0)\}$. The March C− test can be transformed into a transparent test (TMarch C−) with the transformation rules shown above. Thus TMarch C− is $\{\Uparrow (rD_a, wD_{\bar{a}}); \Uparrow (rD_{\bar{a}}, wD_a); \Downarrow (rD_a, wD_{\bar{a}}); \Downarrow (rD_{\bar{a}}, wD_a); \updownarrow (rD_a)\}$ with $a=0$. The fault coverage of TMarch C− is the same as that of March C−, which is shown in [11, 12]. Also, the signature prediction algorithm of TMarch C− is $\{\Uparrow (rD_a); \Uparrow (rD_{\bar{a}}); \Downarrow (rD_a); \Downarrow (rD_{\bar{a}}); \updownarrow (rD_a)\}$.

Similarly, a conventional word-oriented march test can be transformed into a transparent word-oriented march test with applying the transformation rules to all the bits of each word. A conventional word-oriented march test can be obtained by running the corresponding bit-oriented march test with different data backgrounds (a Read or Write operation for a word-oriented memory involves reading or writing an entire word of data, called data background [4]). For example, if March C− is used to test a memory with 4-bit words, the word-oriented March C− test will be: $\{\updownarrow (wX); \Uparrow (rX, w\overline{X}); \Uparrow (r\overline{X}, wX); \Downarrow (rX, w\overline{X}); \Downarrow (r\overline{X}, wX); \updownarrow (rX)\}$ for $X=\{0000, 0101, 0011\}$. Thus the test consists of the following three parts:

T1:$\{\updownarrow (w0000); \Uparrow (r0000, w1111); \Uparrow (r1111, w0000); \Downarrow (r0000, w1111); \Downarrow (r1111, w0000); \updownarrow (r0000)\}$;

T2:$\{\updownarrow w0101; \Uparrow (r0101, w1010); \Uparrow (r1010, w0101); \Downarrow (r0101, w1010); \Downarrow (r1010, w0101); \updownarrow (r0101)\}$;

T3:$\{\updownarrow w0011; \Uparrow (r0011, w1100); \Uparrow (r1100, w0011); \Downarrow (r0011, w1100); \Downarrow (r1100, w0011); \updownarrow (r0011)\}$;

This test can be transformed into a transparent test with executing the transformation rules shown above on each bit of a word [12], and the transparent test can be otained as follows:

T1':$\{\Uparrow(rD_{a_0}, wD_{\bar{a}_0}); \Uparrow(rD_{\bar{a}_0}, wD_{a_0}); \Downarrow(rD_{a_0}, wD_{\bar{a}_0}); \Downarrow(rD_{\bar{a}_0}, wD_{a_0}); \updownarrow(rD_{a_0})\}$;

T2':$\{\updownarrow (wD_{a_1}); \Uparrow (rD_{a_1}, wD_{\bar{a}_1}); \Uparrow (rD_{\bar{a}_1}, wD_{a_1}); \Downarrow (rD_{a_1}, wD_{\bar{a}_1}); \Downarrow (rD_{\bar{a}_1}, wD_{a_1}); \updownarrow (rD_{a_1})\}$;

T2':$\{\updownarrow (wD_{a_2}); \Uparrow (rD_{a_2}, wD_{\bar{a}_2}); \Uparrow (rD_{\bar{a}_2}, wD_{a_2}); \Downarrow (rD_{a_2}, wD_{\bar{a}_2}); \Downarrow (rD_{\bar{a}_2}, wD_{a_2}); \updownarrow (rD_{a_2})\}$;

T4':$\{\updownarrow (wD_{a_0})\}$;

where $a_0=\{0000\}$, $a_1=\{0101\}$, and $a_2=\{0011\}$. Note that the additional T4' test is used for restoring the initial data of the memory under test.

## 4 Proposed Transparent March Tests

The transparent test approach presented in this paper is applied in idle state of systems. Reducing the test time thus is very important for avoiding the interrupt of testing. As discussed in Sec. 3, however, a transparent word-oriented march test are directly obtained by repeatly executing the corresponding bit-oriented march test on each bit of a word. This transformation does not generate an efficient transparent word-oriented march test. In this section we propose a transformation algorithm for transforming a bit-oriented march test into a time-efficiency transparent word-oriented march test.

The proposed transparent word-oriented march transformation algorithm (TWM_TA) is described in Algorithm 1. When a bit-oriented march test (BMarch) is given, TWM_TA first replaces $w0$ or $w1$ of the BMarch with $w\mathbf{0}$ or $w\mathbf{1}$, where $\mathbf{0}$ and $\mathbf{1}$ denote the all-0 and all-1 data backgrounds. Also, the new march test with solid data backgrounds is called SMarch. If the last operation of SMarch is a Write operation, then added a Read operation in the end of SMarch. Subsequently, TWM_TA transforms the SMarch into a transparent SMarch (TSMarch) with the transformation rules for bit-oriented march tests (described in Sec. 3) by regarding the SMarch as a bit-oriented march test. When the SMarch is executed, if the expected data for the last Write operation of every word is equal to the inverse of the initial data of the word, then added an additional transparent test ATMarch=$\{\updownarrow(wD_{a_i}, wD_{\bar{a}_i}, rD_{\bar{a}_i}, wD_{a_i}, rD_{a_i}); \updownarrow (wD_{\bar{a}_0})\}$. On the contrary, if the expected data for the last Write operation of every word is equal to the initial data of the word, then added an additional transparent test ATMarch= $\{\updownarrow (wD_{a_i}, wD_{\bar{a}_i}, rD_{\bar{a}_i}, wD_{a_i}, rD_{a_i}); \updownarrow (wD_{a_0})\}$. Note that the $a_i$ in ATMarch is a $B$-bit binary data and $a_i=\sum_{j=0}^{B-1} x_j 2^j$, where $x_j=1$ or $x_j=0$ if $\lfloor j/i \rfloor$ is even or odd for $i=1,2,\ldots,\lceil \log_2 B \rceil$, for a memory with $B$-bit words.

For example, if a memory with 8-bit words is tested, then three data patterns need to be applied to ATMarch, i.e., $i=\{1, 2, 3\}$. According to the description above, the ATMarch must be executed with the following $a_i$, $a_1=\{01010101\}$ since $\lfloor j/i \rfloor=\{7,6,5,4,3,2,1,0\}$ for $i=1$; $a_2=\{00110011\}$ since $\lfloor j/i \rfloor=\{3,3,2,2,1,1,0,0\}$ for $i=2$; $a_3=\{00001111\}$ since $\lfloor j/i \rfloor=\{1,1,1,1,0,0,0,0\}$ for $i=3$. Thus ATMarch=$\{\updownarrow(wD_{a_1}, wD_{\bar{a}_1}, rD_{\bar{a}_1}, wD_{a_1}, rD_{a_1}); \updownarrow(wD_{a_2}, wD_{\bar{a}_2}, rD_{\bar{a}_2}, wD_{a_2}, rD_{a_2}); \updownarrow(wD_{a_3}, wD_{\bar{a}_3}, rD_{\bar{a}_3}, wD_{a_3}, rD_{a_3}); \updownarrow(wD_{a_0} \text{ or } wD_{\bar{a}_0})\}$. Assume that the content of the word is $D=\{d_7 d_6 d_5 d_4 d_3 d_2 d_1 d_0\}$ after the TSMarch is performed. Table 1 shows the content of the word when the first three march elements of ATMarch are executed.

An example is given to explain the proposed transparent march test scheme further. Consider a bit-oriented March U [15] :$\{\updownarrow (w0); \Uparrow (r0, w1, r1, w0); \Uparrow (r0, w1); \Downarrow (r1, w0, r0, w1); \Downarrow (r1, w0)\}$. The March U is first transformed into a test with solid backgrounds, called SMarch U.

**Algorithm 1** TWM_TA

**Require:** BMarch—a given bit-oriented march test
    TBMarch: the transparent BMarch converted from BMarch
    **0** or **1**: all-0 data or all-1 data
    SMarch: BMarch with solid data backgrounds (all-0 or all-1)
    TSMarch: the corresponding transparent test of SMarch
    ATMarch: an added transparent march test
    TWMarch: transparent word-oriented march test
    Word[$j$]: the content of the $j$th word
    $a_i$: a $B$-bit binary data and $a_i = \sum_{j=0}^{B-1} x_j 2^j$, where $x_j = 1$
    or $x_j = 0$ if $\lfloor j/i \rfloor$ is even or odd for $i = 1, 2, \ldots, \lceil \log_2 B \rceil$,
    for a memory with $B$-bit words.
    $a_0$: a all-0 data
    **if** BMarch = {ø} **then**
      Abort
    **while** BMarch ≠ {ø} **do**
      **for all** Test Operations of BMarch **do**
        Replace $w0$ or $w1$ with $w\mathbf{0}$ or $w\mathbf{1}$, respectively
        Replace $r0$ or $r1$ with $r\mathbf{0}$ or $r\mathbf{1}$, respectively
      **if** The last operation of SMarch is a Write operation **then**
        Added a Read operation in the end of SMarch
        Transform SMarch into TSMarch according to the transformation rules described in Sec. 3 [11, 12]
      **else**
        Transform SMarch into TSMarch according to the transformation rules described in Sec. 3 [11, 12]
      **if** Word[$j$]≠the initial content after executing the TSMarch **then**
        ATMarch= {⇕ $(wD_{a_i}, wD_{\bar{a}_i}, rD_{\bar{a}_i}, wD_{a_i}, rD_{a_i})$; ⇕ $(wD_{a_0})$}
      **else**
        ATMarch= {⇕ $(wD_{a_i}, wD_{\bar{a}_i}, rD_{\bar{a}_i}, wD_{a_i}, rD_{a_i})$; ⇕ $(wD_{\bar{a}_0})$}
    Return TWMarch={TSMarch; ATMarch}
    Return Signature prediction test by removing the Write operations in TWMarch

Then SMarch U is {⇕ $w\mathbf{0}$; ⇑ $(r\mathbf{0}, w\mathbf{1}, r\mathbf{1}, w\mathbf{0})$; ⇑ $(r\mathbf{0}, w\mathbf{1})$; ⇓ $(r\mathbf{1}, w\mathbf{0}, r\mathbf{0}, w\mathbf{1})$; ⇓ $(r\mathbf{1}, w\mathbf{0})$}. Because the last operation of SMarch U is a Write operation, an additional Read operation (⇕ $(r\mathbf{0})$) is added in the end of the SMarch U. Assume that a memory with 8-bit words is tested. According to the transformation rules described in [11, 12], TSMach U will be {⇑ $(rD_{a_0}, wD_{\bar{a}_0}, r_{D_{\bar{a}_0}}, wD_{a_0})$; ⇑ $(rD_{a_0}, wD_{\bar{a}_0})$; ⇓ $(rD_{\bar{a}_0}, wD_{a_0}, rD_{a_0}, wD_{\bar{a}_0})$; ⇓ $(rD_{\bar{a}_0}, wD_{a_0})$; ⇕ $(rD_{a_0})$}, where $a_0$={00000000}. After TSMarch is excuted, the content of each word is equal to the initial content of the word. Therefore, the added ATMarch is {⇕$(wD_{a_1}, wD_{\bar{a}_1}, rD_{\bar{a}_1}, wD_{a_1}, rD_{a_1})$; ⇕$(wD_{a_2}, wD_{\bar{a}_2}, rD_{\bar{a}_2}, wD_{a_2}, rD_{a_2})$; ⇕$(wD_{a_3}, wD_{\bar{a}_3}, rD_{\bar{a}_3}, wD_{a_3}, rD_{a_3})$; ⇕$(wD_{a_0})$} where $a_1$={01010101}, $a_1$={00110011}, and $a_1$={00001111}. The complexity of the transformed transparent word-oriented March U is $29N$ for testing a memory with 8-bit words. Also, time complexity of the corresponding signature prediction algorithm is $14N$, which can be obtained by removing the Write operations. Formal analysis of time complexity of the proposed transparent test scheme will be discussed in Sec. 5.

Table 1: The content of the word when the first three march elements of ATMarch are executed.

| Test operations | $d_7 d_6 d_5 d_4 d_3 d_2 d_1 d_0$ |
|---|---|
| $wD_{a_1}$ | $d_7 \bar{d_6} d_5 \bar{d_4} d_3 \bar{d_2} d_1 \bar{d_0}$ |
| $wD_{\bar{a}_1}, rD_{\bar{a}_1}$ | $\bar{d_7} d_6 \bar{d_5} d_4 \bar{d_3} d_2 \bar{d_1} d_0$ |
| $wD_{a_1}, rD_{a_1}$ | $d_7 \bar{d_6} d_5 \bar{d_4} d_3 \bar{d_2} d_1 \bar{d_0}$ |
| $wD_{a_2}$ | $d_7 \bar{d_6} \bar{d_5} d_4 d_3 \bar{d_2} \bar{d_1} d_0$ |
| $wD_{\bar{a}_2}, rD_{\bar{a}_2}$ | $\bar{d_7} d_6 d_5 \bar{d_4} \bar{d_3} d_2 d_1 \bar{d_0}$ |
| $wD_{a_2}, rD_{a_2}$ | $d_7 \bar{d_6} \bar{d_5} d_4 d_3 \bar{d_2} \bar{d_1} d_0$ |
| $wD_{a_3}$ | $d_7 d_6 d_5 d_4 \bar{d_3} \bar{d_2} \bar{d_1} \bar{d_0}$ |
| $wD_{\bar{a}_3}, rD_{\bar{a}_3}$ | $\bar{d_7} \bar{d_6} \bar{d_5} \bar{d_4} d_3 d_2 d_1 d_0$ |
| $wD_{a_3}, rD_{a_3}$ | $d_7 d_6 d_5 d_4 \bar{d_3} \bar{d_2} \bar{d_1} \bar{d_0}$ |

## 5 Fault Coverage Analysis and Comparison

The fault coverage analysis of the transparent word-oriented march tests transformed with the proposed TWM_TA is discussed. Fault coverage of the faults described in Sec. 2 are considered. For a given bit-oriented march test, we will show that the fault coverage of the transparent word-oriented march test transformed with TWM_TA is the same as the fault coverage of the corresponding word-oriented march test. Apparently, if a bit-oriented march test can detect the SAFs and TFs, the corresponding transformed transparent march test by TWM_TA also can detect the SAFs and TFs, since the TSMarch is obtained with transforming the bit-oriented march test directly.

Subsequently, the fault coverage analysis of CFs is discussed. Figure 1(a) shows all possible fault-free states of any two arbitrary cells in a bit-oriented memory. Assume that the state of the two cells is $(D_i D_j)$ after the initialization march element of a march test. Also, let $i$ be the lower address and $j$ be the higher one without loss of generality. If a bit-oriented march test can detect 100% CFs (CFin, CFid, and CFst), then test operations of the march test must excite any two arbitrary cells to undergo all the states as shown in Fig. 1(a). For example, the March C− has been shown that it can detect 100% CFs [14]. If the March C− is executed in $(D_i D_j)$, the two cells undergo all the states with the sequence 1,2,3,…,18. The TWM_TA first transforms a BMarch into a corresponding SMarch. Apparently, the SMarch also can detect 100% inter-word CFs if the BMarch can detect 100% CFs. Thus the $(D_i D_j)$ of Fig. 1 can be replaced with any two arbitrary words. Also, SMarch can undergo the states with the sequence which is the same as that of BMarch. Therefore, TSMarch also can detect 100% inter-word CFs, since fault coverage of TSMach is unchanged after the transformation rules reported in [12] are executed.

In a word-oriented memory, all bits of a word perform the Read or Write operation concurrently. Figure 1(b) shows all possible fault-free states of any two bits within a word. A word-oriented march test which can cover 100% intra-word CFs (CFid, CFin, and CFst) makes any two bits within a word undergo the following state con-ditions: $(d_i \to \bar{d}_i; d_j)$, $(d_i \to \bar{d}_i; \bar{d}_j)$, $(\bar{d}_i \to d_i; d_j)$, and $(\bar{d}_i \to d_i; \bar{d}_j)$, where $(x \to \bar{x}; y)$ denotes a Write operation writing the data $(\bar{x}y)$ into two bits within a word and the original data of the two bits is $(xy)$, and subsequently a Read operation reads the expected data $(\bar{x}y)$. As Fig. 1(a) shows, if the $D_i$ and $D_j$ is replaced with two arbitrary words, then the two state conditions, $(d_i \to \bar{d}_i; d_j)$ and $(\bar{d}_i \to d_i; d_j)$, can be checked when the SMarch is executed. Finally, the non-transparent part of ATMarch is AMarch=$\{\Updownarrow (wa_1, w\bar{a}_1, r\bar{a}_1, wa_1, ra_1)\}$ for $a_1 = \{01\}$ is executed, where a memory with two-bit words is assumed without loss of generality. Assume that the memory state is all-0 when the SMarch is completed. That is, the AMarch executes the following operations for each word: $\{(wd_i\bar{d}_j, w\bar{d}_i d_j, r\bar{d}_i d_j, wd_i\bar{d}_j, rd_i\bar{d}_j)\}$. The state condition $(d_i \to \bar{d}_i; d_j)$ is checked when the $r\bar{d}_i d_j$ is executed; and the condition $(\bar{d}_i \to d_i; \bar{d}_j)$ is also checked when the $rd_i\bar{d}_j$ is performed. Therefore, the four state conditions for intra-word CFs detection are checked when SMarch and AMarch are performed if the corresponding bit-oriented march test can detect 100% CFs. According to the test coverage theorem in [12], we conclude that the transformed transparent word-oriented march test (TWMarch=TSMarch+ATMarch) can preserve the same fault coverage of inter-word and intra-word CFs of the corresponding nontransparent word-oriented march test (SMarch+AMarch).

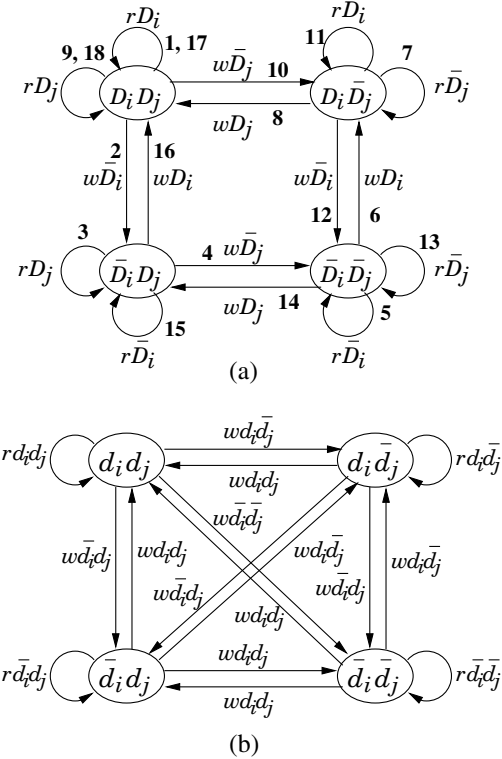

Figure 1: (a) All states of two arbitrary words when TSMarch is executed. (b) All states of two arbitrary bits within a word.

In the sequel, we analyze time complexity of the transparent word-oriented march tests converted by the TWM_TA and compare the time complexity with other transparent test schemes. Consider an $N \times B$-bit memory and a bit-oriented march (BMarch) test with $P$ Read/Write operations in which there are $Q$ Read operations. Without loss of generality, we assume that BMarch has an initialization Write operation, the first operation is Read operation in each march element, and the last operation is a Read operation. Also, we assume that $B$ is power of 2. As Algorithm 1 shows, time complexity of the transparent word-oriented march test (TCM) transformed with TWM_TA with respect to the bit-oriented march test can be calculated as follows:

$$\text{TCM} = (P + 5\log_2 B)N.$$

Also, time complexity of the corresponding signature prediction test (TCP) can be expressed as follows:

$$\text{TCP} = (Q + 2\log_2 B)N.$$

We compare the proposed transparent march tests with the transparent march tests reported in [12, 13], since the two works also discuss the word-oriented transparent tests. As described in Sec. 3, the transparent test scheme reported in [12] transforms the bit-oriented march tests into transparent word-oriented march tests with executing the transformation in each bit of a word. Therefore, the TCM and TCP of the transparent word oriented march test converted with the scheme reported in [12] can be calculated as

$$\text{TCM} = P(\log_2 B + 1)N.$$
$$\text{TCP} = Q(\log_2 B + 1)N.$$

The TCM of the transparent online march test (TOMT) reported in [13] is $(4 + 8B)N$. TOMT is a online test which does not need the signature prediction test, i.e., TCP=0. Table 2 summarizes the comparison of TCM and TCP for the three transparent test schemes discussed above.

Table 2: Comparison of different transparent test schemes.

|  | TCM | TCP |
|---|---|---|
| Scheme 1 [12] | $P(\log_2 B + 1)N$ | $Q(\log_2 B + 1)N$ |
| Scheme 2 [13] | $(4 + 8B)N$ | No |
| This work | $(P + 5\log_2 B)N$ | $(Q + 2\log_2 B)N$ |

As Table 2 shows, the proposed transparent word-oriented test scheme has better time complexity compared with Scheme 1 [12]. Also, the proposed scheme has shorter time compared with Scheme 2 [13] for most cases. For example, if March C− is transformed into a transparent word-oriented March C− for testing an $N \times 32$-bit memory, then test complexity (TCP+TCM) of Scheme 1 [12], Scheme 2 [13], and the proposed scheme is $90N$, $260N$, and $50N$, respectively. That is, time complexity of the transparent word-oriented test transformed by the proposed scheme is only about 56% or 19% time complexity of the transparent word-oriented test converted by Scheme 1 or Scheme 2, respectively. Table 3 summarizes the test complexity of the three schemes for a memory with respect to different word sizes and test algorithms. As the table shows, the transparent test scheme proposed in this paper has smaller time complexity compared with the other two works. Furthermore, time complexity of the proposed transparent test scheme is only slightly related to the bit-oriented test. However, time complexity of the transparent test scheme reported in [12] is heavily related to the bit-oriented test.

Table 3: Comparison of time complexity for different word sizes.

| Test | Word Size | [12] | [13] | This work |
|------|-----------|------|------|-----------|
| March C− | 16 bits | $75N$ | $132N$ | $43N$ |
|  | 32 bits | $90N$ | $260N$ | $50N$ |
|  | 64 bits | $105N$ | $516N$ | $57N$ |
|  | 128 bits | $120N$ | $1028N$ | $64N$ |
| March U | 16 bits | $95N$ | $132N$ | $47N$ |
|  | 32 bits | $114N$ | $260N$ | $54N$ |
|  | 64 bits | $133N$ | $516N$ | $61N$ |
|  | 128 bits | $152N$ | $1028N$ | $68N$ |

## 6 Conclusions

This paper presents a systematic algorithm for transforming a bit-oriented march test into a transparent word-oriented march test. We have also shown that fault coverage of the transformed transparent word-oriented march test is the same as that of the corresponding nontransparent word-oriented march test. Consider an $N \times B$-bit memory and a bit-oriented march test with $P$ Read/Write operations in which there are Q Read operations. Time complexity of the transparent word-oriented march test converted by the proposed algorithm is $(P+5\log_2 B)N$. Also, time complexity of the corresponding signature prediction test is $(Q+2\log_2 B)N$. The transformed transparent march test has shorter test time compared with that proposed in previous works. For example, if a memory with 32-bit words is tested with March C−, the time complexity of the transparent test transformed by the proposed scheme is only about 56% or 19% time complexity of the transparent test converted by the scheme [12] or [13]. Moreover, compared with the scheme [12], time complexity of the proposed transparent test scheme is slightly related to the corresponding bit-oriented march tests.

## Acknowledgement

This work was supported in part by the National Science Council, R.O.C., under Contract NSC 92-2218-E-008-005.